\documentclass[%
 reprint,
 amsmath,amssymb,
 aps,
prb,
]{revtex4-1}

\usepackage{graphicx}% Include figure files
\usepackage{dcolumn}% Align table columns on decimal point
\usepackage{bm}% bold math
\usepackage{here}
\begin{document}

\preprint{APS/123-QED}

\title{Theoretical study on the correlation between the spin fluctuation  and $T_c$ in the isovalent-doped 1111 iron-based superconductors}% Force line breaks with \\
%\thanks{A footnote to the article title}%

\author{
Hayato Arai$^1$, Hidetomo Usui$^2$, Katsuhiro Suzuki$^2$, Yuki Fuseya$^1$, and Kazuhiko Kuroki$^2$}
\affiliation{$^1$ Department of Engineering Science, 
University of Electro-Communications, Chofu, Tokyo 182-8585, Japan}
\affiliation{$^2$ Department of Physics, 
Osaka University, Toyonaka, Osaka 560-0043, Japan}

\date{\today}% It is always \today, today,
             %  but any date may be explicitly specified

\begin{abstract}
Motivated by recent experiments on isovalent-doped 1111 iron-based superconductors $\rm LaFeAs_{1-{\it x}}P_{\it x}O_{1-{\it y}}F_{\it y}$ and the theoretical study that followed, 
we investigate, within the five orbital model, the correlation between the spin fluctuation and the superconducting transition temperature, which exhibits a double dome feature upon varying the Fe-As-Fe bond angle. Around the first dome with higher $T_c$, the low energy spin fluctuation and $T_c$ are not tightly correlated because the finite energy spin fluctuation also contributes to superconductivity. On the other hand, the strength of the low-energy spin fluctuation originating from the $d_{xz/yz}$ orbital is correlated with $T_c$ in the second dome with lower $T_c$. These calculation results are consistent with recent NMR study, and hence strongly suggest that the pairing in the iron-based superconductors is predominantly caused by the multi-orbital spin fluctuation.

\begin{description}
%\item[Usage]
%Secondary publications and information retrieval purposes.
\item[PACS numbers]
74.20.-z, 74.70.Xa, 74.25.nj%May be entered using the \verb+\pacs{#1}+ command.
%\item[Structure]
%You may use the \texttt{description} environment to structure your abstract;\onlinecite
%use the optional argument of the \verb+\item+ command to give the category of each item. 
\end{description}
\end{abstract}

\pacs{Valid PACS appear here}% PACS, the Physics and Astronomy
                             % Classification Scheme.
%\keywords{Suggested keywords}%Use showkeys class option if keyword
                              %display desired
\maketitle

%\tableofcontents

\section{\label{sec:level1}INTRODUCTION}
It is now well known that the superconducting transition temperature ($T_c$) of the iron-based superconductors can vary, in some cases largely, by substituting some elements with others. In particular cases, the $T_c$ variation upon element substitution has been found to be non-monotonic.
These observations have sometimes lead to a speculation that the Cooper pairing in the iron-based superconductors may involve multiple pairing mechanisms.
One of the recent important example is the hydrogen-doped 1111 system\cite{Iimura}. Namely, large amount of electrons can be doped by substituting O with H in {\it Ln}FeAsO with ({\it Ln}=Gd, Sm, Ce, La).
Surprisingly, it has been shown that the superconductivity appears even up to 40\% of electron doping, and particularly in LaFeAs(O,H)\cite{Iimura} and SmFe(As,P)(O,H)\cite{Matsuishi} the phase diagram exhibits a double-dome structure against the electron doping.
Some of the present authors of the present paper have clarified the origin of this double dome phase diagram\cite{Suzuki}. Namely, while the first $T_c$ dome is  mainly due to the Fermi surface nesting, the second dome originates from the spin fluctuation enhanced by a peculiar real space motion of electrons within the  $d_{xy}$ orbital, despite the degraded nesting.

Another example was recently found in an isovalent doping 1111 system, where As is partially substituted with P in {\it Ln}FeAs(O,F) ({\it Ln}=Nd, Ce, La).
It is known that increasing the phosphorous content enlarges the Fe-{\it Pn}-Fe ({\it Pn}=pnictogen) bond angle (or reduces the pnictogen height). Therefore, it was expected, from the empirical $T_c$ trend found by Lee et al\cite{Lee}, that $T_c$ monotonically decreases  by increasing the phosphorous content\cite{Lee,KurokiPRB,UsuiPRB}. However, it has recently been revealed that another local maximum of $T_c$ (we will call this the second $T_c$ dome) exists in the high phosphorous content regime\cite{Miyasaka,Mukuda,Lai}.
Moreover, for the non-fluorine-doped system, the antiferromagnetic phase sandwiched by the two superconducting phases  were found in the finite P doping region\cite{Ishida2,Mukuda2,Lai}. In order to investigate this problem theoretically, some of the authors of the present paper have recently studied the correlation between the Fe-{\it Pn}-Fe bond angle and the spin-fluctuation-mediated superconductivity\cite{UsuiSciRep}.
There, it was shown that the eigenvalue of the linearized Eliashberg's equation, $\lambda$, which can be considered as a measure for the superconducting $T_c$, and also the Stoner factor of antiferromagnetism, vary non-monotonically as functions of the bond angle, indeed exhibiting a double dome structure consistent with the experiments. They concluded that in the small bond angle region (small P concentration regime), the spin fluctuations originating from the $d_{xy}$ and $d_{xz/yz}$ orbitals both contribute to the superconductivity, while in the large bond angle region (large P concentration), the $d_{xy}$ orbital has small contribution to superconductivity, and the nesting of the $d_{xz/yz}$ portion of the Fermi surface is the main origin of the superconductivity.

Experimentally, the magnitude of the low energy spin fluctuation is often probed by the NMR experiment. Namely, $1/T_1 T$ measured in the NMR experiment is proportional to the slope of the imaginary part of the dynamical spin susceptibility  taken in the $\omega \rightarrow 0$ limit. Therefore, the correlation between the superconducting $T_c$ and the development of $1/T_1 T$ upon lowering the temperature in the normal state has often been discussed in the context of determining whether the Cooper pairing is mediated by the spin fluctuation or not. In fact, in some cases, the superconducting $T_c$ and the development of $1/T_1 T$ at low temperature is found to be correlated\cite{11press,122Co,122AsP}, suggesting the importance of the spin fluctuation played in the pairing mechanism. 

\begin{figure}[tbp]
\begin{center}
\includegraphics[scale=0.36]{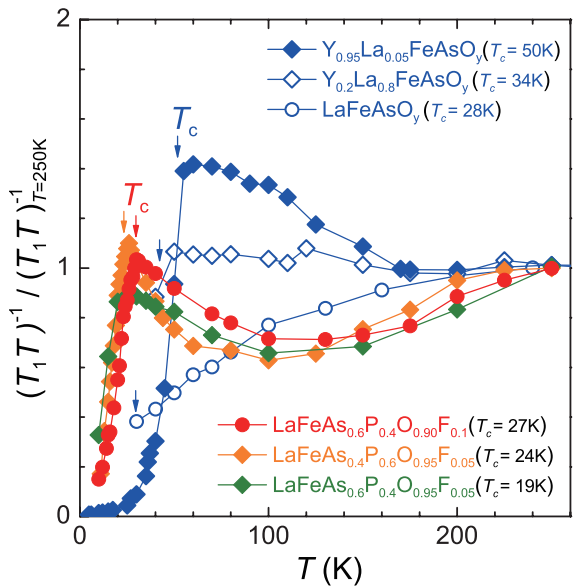}
\caption{The NMR experiment result of $1/T_1T$ against $T$ for 
LaFeAs$_{1-{\it x}}$P$_{\it x}$O$_{1-{\it y}}$F$_{\it y}$ and Y$_{\it z}$La$_{1-{\it z}}$FeAsO$_{\it y}$. Reprinted from ref.[\onlinecite{Mukuda}]. }
\label{fig:zero}
\end{center}
\end{figure}

As for LaFe(As,P)(O,F) mentioned above, the correlation between $T_c$ and $1/T_1 T$ has been investigated in ref.[\onlinecite{Mukuda}] as shown in Fig.\ref{fig:zero}. It was found that $T_c$ at finite phosphorous content is correlated with the development of $1/T_1 T$ upon lowering the temperature, indicating that the low energy spin fluctuation is responsible for the non-monotonic $T_c$ dependence with phosphorous doping. 
On the other hand, it was also shown in the same paper that in a 1111 material La$_{0.05}$Y$_{0.95}$FeAsO$_{1-y}$, in which $T_c$ reaches a very high $T_c$ of $\sim 50$K,  the development of $1/T_1 T$ tends to be saturated in the low temperature regime just above $T_c$ (Fig.\ref{fig:zero}). Comparing this result to that for LaFe(As,P)(O,F) suggests that the $T_c$ and the development of the low energy spin fluctuation is not necessarily correlated, indicating that factors other than the {\it low energy} spin fluctuation is playing a role in the enhancement of the superconductivity.
In fact, the observation that $T_c$ is not necessarily correlated with $1/T_1 T$ has already been recognized shortly after the discovery of the iron-based superconductors. Namely, in ref.[\onlinecite{Ishida}], it was pointed out that $T_c$ is not largely affected by doping electrons into LaFeAs(O,F), while the development of $1/T_1 T$ at low temperatures is suppressed prominently. Therefore, it seems that $1/T_1 T$ is sometimes correlated with $T_c$, while in other cases not. It is interesting to see whether these experimental observations can be understood within the framework of the spin-fluctuation mediated superconductivity consistently. The doping dependence of $1/T_1 T$  has indeed been investigated by Ikeda in the early days\cite{Ikeda1,Ikeda2}, but the non-monotonic $T_c$ behavior in LaFe(As,P)(O,F) was not known at that time.

Given the above mentioned background, in the present study, we theoretically study the correlation between the Fe-{\it Pn}-Fe bond angle, the strength of the $\omega\sim 0$ or finite energy spin fluctuation, and the superconductivity mediated by spin fluctuation. We conclude that the existing experimental results can indeed be understood within this framework, where the key point is the contribution to superconductivity coming from finite energy spin fluctuation that has small contribution to NMR $1/T_1T$.

\section{\label{sec:level2}Formulation}
\subsection{Model construction}
We first perform first principles band calculation of $\rm LaFeAsO$ using 
the WIEN2k package\cite{Wien2k}. We adopt hypothetical lattice structures of LaFeAsO, where the Fe-As-Fe bond angle $\alpha$ is varied, while fixing the Fe-As bond length. From the first principles band calculation, using the maximally localized Wannier orbitals\cite{MaxLoc, wannier90, w2w} and following ref.[\onlinecite{KurokiPRL}], we construct the  five orbital tight-binding model in the unfolded Brillouin zone in the following form 
\begin{eqnarray*}
H_0=\sum_{i,j}{\sum_{\mu,\nu}{\sum_{\sigma}{t_{ij}^{\mu \nu} c_{i \mu \sigma}^{\dagger}c_{j \nu \sigma}}}}
 \end{eqnarray*},
where $c_{i,j \mu,\nu \sigma}^{\dagger}$ creates an electron with spin $\sigma$ on the $\mu,\nu$ th orbital at site $i,j$, and $t_{ij}^{\mu \nu}$ are the hopping integrals.
We define the band filling $n$ as the number of electrons per number of sites (e.g., $n=10$ for full filling). On top of the non-interacting part $H_0$, we take into account the electron-electron interaction  in the following form : 
\begin{eqnarray*}
H_1= \sum_i [\sum_{\mu} U_{\mu} n_{i \mu \uparrow} n_{i \mu \downarrow} + \sum_{\mu > \nu}\sum_{\sigma \sigma'} U'_{\mu \nu} n_{i \mu \sigma} n_{i \mu \sigma'} \\
 - \sum_{\mu \neq \nu} J_{\mu \nu} {\bold S_{\rm i \mu} \cdot \bold S_{\rm i \nu}} + \sum_{\mu \neq \nu} J'_{\mu \nu} c_{i \uparrow}^{\mu \dagger} c_{i \downarrow}^{\mu \dagger} c_{i \downarrow}^{\nu} c_{i \uparrow}^{\nu}],
 \end{eqnarray*}
where the standard interaction terms that comprise the intra-orbital Coulomb interaction $U$, the inter-orbital Coulomb interaction $U'$, the Hund's coupling $J$, and the pair-hopping $J'$ are considered.
The magnitude of the interactions are taken to be dependent on the orbital.

\subsection{FLuctuation EXchange approximation (FLEX)}
In the present study, we adopt The fluctuation exchange (FLEX) approximation\cite{Bickers,Dahm} to the multi-orbital Hubbard model. 
The renormalized Green's function $G$ is given as 
\begin{eqnarray*}
G = (G_0^{-1} -\Sigma)^{-1}
 \end{eqnarray*}
where $G_0$ is the Green's function of the non-interacting electrons and $\Sigma$ is the self energy. In FLEX, bubble and ladder type diagrams are considered in the self energy calculation, which is determined self-consistently in the Baym-Kadanoff sense in order to have the conservation laws satisfied.
The spin and charge susceptibilities matrices $\chi_s(q)$ and $\chi_c(q)$, respectively, are described  in this approximation by the following equations, 
\begin{eqnarray*}
\chi_s(q) = \frac{\chi_0 (q)}{1 - S \chi_0 (q)} \\
\chi_c(q) = \frac{\chi_0 (q)}{1 + C \chi_0 (q)}
 \end{eqnarray*}
where $S$ and $C$ are orbital dependent interaction vertex matrices.
 $q \equiv ({\bf q},i\omega_n)$, where ${\bf q}$ is the wave vector and $i \omega_n \equiv (2n + 1)\pi k_{B}T$ is the Matsubara frequency.
The irreducible susceptibility matrix elements $\chi^{l_1l_2l_3l_4}_0(q)$ is given using the renormalized Green's function as 
\begin{eqnarray*}
\chi^{l_1l_2l_3l_4}_0(q) = \sum_{k}{G_{l_1 l_3} (k+q)G_{l_4 l_2} (k)},
\end{eqnarray*}
where $l_i$ ($i=1,2,\cdots,5$) are orbital indices. 

After $\chi_s(q)$ is obtained on the imaginary frequency axis,  we perform analytic continuation exploiting the $\rm Pad \acute{e}$ approximation to obtain the spin susceptibility on the real frequency axis.
We define quantities that measure the strength of the intra orbital 
spin fluctuation as 
 \begin{eqnarray*}
\sum_{\bf k} {{\rm Im} \chi_s ^{\mu}({\bf k},\omega)} \equiv \Gamma^{\mu}(\omega) \\
\frac{1}{T_1 T} \approx \lim_{\omega \rightarrow 0}{\sum_{\mu} {\frac{\Gamma^{\mu}(\omega)}{\omega}}} \equiv \sum_{\mu}{\gamma^{\mu}}
 \end{eqnarray*}
Here, the intraorbital spin susceptibility of the $\mu$-th orbital is 
$\chi_s^{\mu}=\chi_s^{\mu\mu\mu\mu}$. $\Gamma^\mu(\omega)$ measures the strength of the spin fluctuation within the $\mu$-th orbital at a certain frequency $\omega$, while $\gamma^\mu$ is a measure of the intraorbital spin fluctuation in the low energy limit. 

 In order to analyze the superconductivity, the singlet pairing interaction is described by the following equation:
  \begin{eqnarray*}
V_{s}(q) = \frac{3}{2}S \chi_s(q)S - \frac{1}{2}C \chi_c(q) C +\frac{1}{2}(S +C).
 \end{eqnarray*}
 Then, the linearized Eliashberg's equation is given as
 \begin{eqnarray*}
\lambda \phi_{l_1 l_4}(k) = -\frac{T}{N} \sum_q {\sum_{l_1 l_2 l_5 l_6} {V_{l_1 l_2 l_3 l_4} (q) G_{l_2 l_5}(k - q)}} \\
\times \phi_{l_5 l_6} (k - q) G_{l_3 l_6} (q - k),
\end{eqnarray*}
where $\phi(q)$ is the gap function. In the present study, the dominant form of the gap function has the $s\pm$ form with some nodes on the electron Fermi surface for large phosphorous content, consistent with previous studies\cite{Mazin,KurokiPRL,Chubukov,Hirschfeld,Ikeda,Daghofer,Thomale,Wang}.
The eigenvalue $\lambda$ monotonically increases with lowering the temperature $T$, and reaches unity at $T=T_c$. We calculate this quantity for a fixed temperature $T>T_c$, so that $\lambda(T)$ is positively correlated with, and thus measure of the  $T_c$. It is known that the spin fluctuation around the wave vector $(\pi,0)/(0,\pi)$ contributes to the $s\pm$ superconductivity, so in the actual calculation of $\Gamma^\mu(\omega)$, the summation over the wave vectors is restricted to the vicinity of $(\pi,0)/(0,\pi)$.

\section{\label{sec:level3}RESULTS}
\subsection{\label{sec:levelA}Bond angle dependence}
\begin{figure}[tbp]
\begin{center}
\includegraphics[scale=0.60]{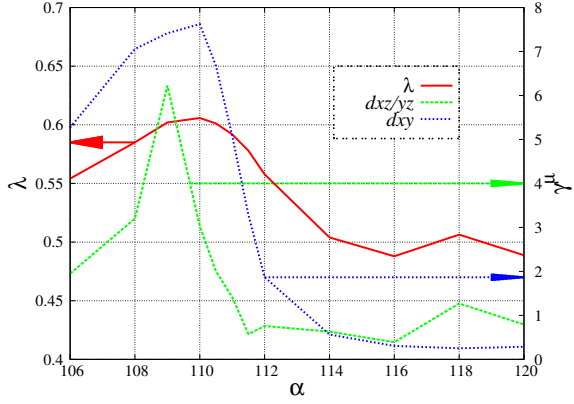}
\caption{Solid red:the eigenvalue of the linearized Eliashberg's equation ;  blue dotted  and green dashed :the low-energy spin-fluctuation $\gamma^{xy}$ and $\gamma^{xz/yz}$, respectively; all against Fe-As-Fe bond angle $\alpha$. We call the local maximum of $\lambda$ around $\alpha = 110$ and $\alpha = 118$ the first and second domes, respectively. }
\label{fig:one}
\end{center}
\end{figure}

In Fig\ref{fig:one}, we show, as functions of the Fe-As-Fe bond angle $\alpha$, the calculated superconducting eigenvalue $\lambda$ as well as the imaginary part  of the low energy derivative of the intraorbital spin susceptibilities $\gamma^{xy}$ and $\gamma^{xz/yz}$. $\lambda$ is calculated for $s$-wave symmetry, and the resulting gap function has the $s\pm$ wave form. In ref.[\onlinecite{UsuiSciRep}], some of the present authors have shown that the variation of $\lambda$ against the bond angle $\alpha$ exhibits a double dome feature as in Fig.\ref{fig:one}, which is consistent with the experimental observations\cite{Miyasaka,Mukuda,Lai}. There, the origin of the double dome structure has been explained as follows: superconductivity is suppressed when the bond angle is increased from $\alpha\sim 110$ because the hole Fermi surface around $(\pi,\pi)$ originating from the $d_{xy}$ orbital vanishes.

However, the spin-fluctuation and $\lambda$ are re-enhanced in the larger bond-angle region (intermediate phosphorous content regime) because the density of states of the inner hole Fermi surface originating from the $d_{xz/yz}$ orbital increases. As the bond angle is further increased (near the phosphide end), the orbital character of the hole Fermi surfaces have the $d_{XZ}$ and $d_{YZ}$ character, where $X$-$Y$ axes are rotated by 45 degrees from $x$-$y$. The electron Fermi surfaces around $(\pi,0)/(0,\pi)$ always have the $d_{xz}$ and $d_{yz}$ character, so that the matching of the orbital character between the electron and the hole Fermi surfaces is not good for too large bond angles, thereby suppressing again the superconductivity.

At first glance, the superconducting eigenvalue and the strength of the low energy spin fluctuation appear to be correlated. However, if we look more closely, we find that this is not necessarily the case around the first dome. Namely, (i) while $\lambda$ barely changes around the bond angle where it is maximized, i.e., $109\leq \alpha\leq111$, the strength of the spin fluctuation in that region strongly varies. At $\alpha=109$, on the left side of the dome, the $d_{xz/yz}$ spin fluctuation is large, but it is strongly suppressed at $\alpha= 111$, on the right side. Similarly, the $d_{xy}$ spin fluctuation is largely suppressed at $\alpha= 111$ compared to $\alpha=110\sim 109$. (ii) On the left and right sides of the first dome, $\alpha=106$ and $\lambda=112$ have almost the same value of $\lambda$,  but the low energy spin fluctuation, especially that for the $d_{xy}$ orbital, is much larger for $\alpha=106$. (iii) At $\alpha=109$, the $d_{xz/yz}$ spin fluctuation exhibits a sharp enhancement, but this does not seem to affect the superconductivity. The low energy spin fluctuation is mainly governed by the details of the Fermi surface nesting condition, so this result indicates that those details and superconductivity are not intimately correlated. 

On the other hand, $\lambda$ around the second dome seems to be correlated with the strength of the $d_{xz/yz}$ low energy spin fluctuation. In this large bond angle regime, the $d_{xy}$ spin fluctuation is always suppressed because of the absence of the $d_{xy}$ Fermi surface around $(\pi,\pi)$\cite{KurokiPRB,UsuiPRB}.

\subsection{Temperature dependence of the low energy spin fluctuation}
\begin{figure}[tbp]
\begin{center}
\includegraphics[scale=0.55]{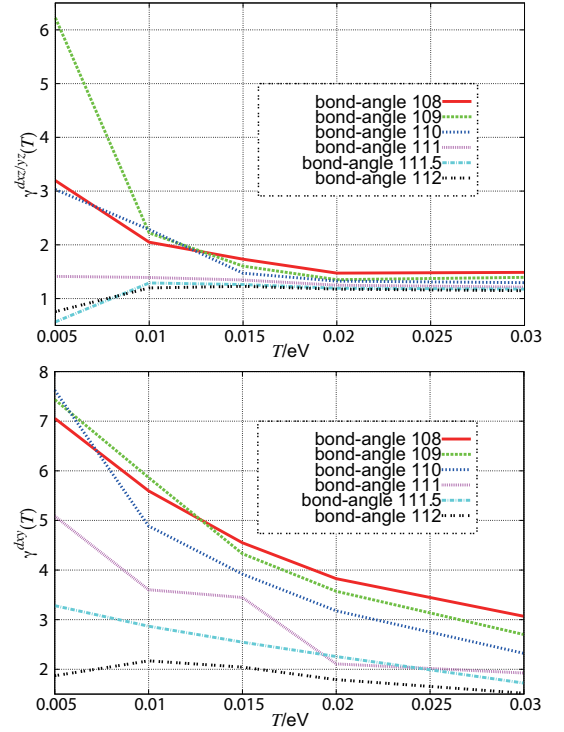}
\caption{The measure for the low-energy spin fluctuation $\gamma^{xz/yz}$ (top) and $\gamma^{xy}$ (bottom) against temperature for various bond angles from 108 to 112 around the first dome.}
\label{fig:two}
\end{center}
\end{figure}

\begin{figure}[tbp]
\begin{center}
\includegraphics[scale=0.50]{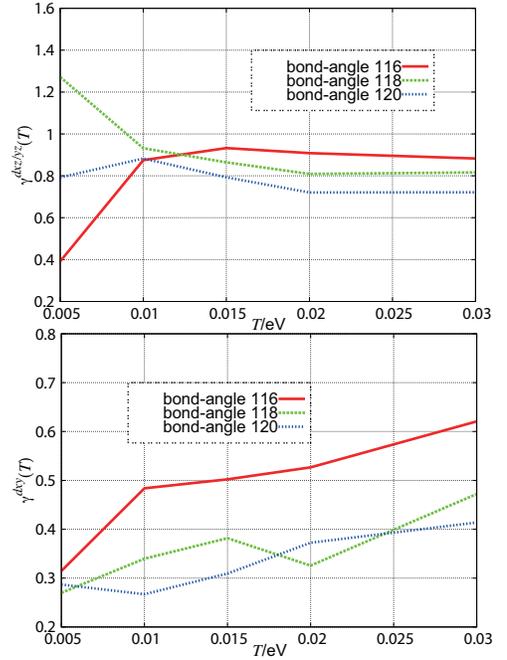}
\caption{Plot similar to Fig.\ref{fig:two} for the bond angles 116, 118, and 120 around the second dome.}
\label{fig:three}
\end{center}
\end{figure}

In order to further investigate the correlation between superconductivity and the low energy spin fluctuation, we now study the temperature dependence of $\gamma^{xy}$ and $\gamma^{xz/yz}$ for various bond angles. In Fig.\ref{fig:two} and Fig.\ref{fig:three}, we show the results for the bond angle within the first and the second domes, respectively. The development of both $\gamma^{xz/yz}$ and $\gamma^{xy}$ is weak even at the bond angle of $\alpha=111$, where the superconducting eigenvalue is close to its maximum value. This further confirms our viewpoint (i) mentioned in the previous section.

As for the second dome,  $\gamma^{xz/yz}$ increases upon lowering the temperature only at $\alpha=118$, where $\lambda$ is locally maximized . In short, the low energy spin fluctuation and superconductivity is not intimately correlated around the first dome, while the superconductivity is correlated with the low energy the $d_{xz/yz}$ spin fluctuation in the second dome.

\subsection{Energy dependence of the spin fluctuation}
\begin{figure}[tbp]
\begin{center}
\includegraphics[scale=0.50]{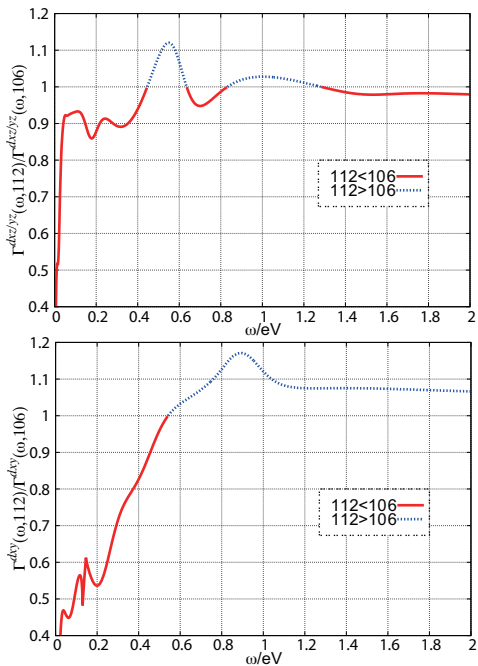}
\caption{The ratio of $\Gamma^{xz/yz}(\omega)$ (top) and $\Gamma^{xy}(\omega)$ (bottom) between bond angle 112 and 106 plotted against ${\omega}$.
Solid red (dotted blue) lines indicate regions where the spin fluctuation for $\alpha=106$ dominates over that for $\alpha=112$.}
\label{fig:four}
\end{center}
\end{figure}

So far we have seen that superconductivity is not tightly correlated with the low energy spin fluctuation around the first dome. The reason for this is expected to be because the finite energy spin fluctuation also contributes to the superconductivity. If the strength of the low energy and finite energy spin fluctuations were correlated, so would superconductivity and the low energy spin fluctuation. Hence, it is expected that the low and finite energy spin fluctuations exhibit different behavior upon varying the bond angle. To see this, we go back to the issue raised as (ii) in section \ref{sec:levelA}, and compare $\Gamma(\omega)$ between the bond angles $\alpha=112$ and $\alpha=106$ by taking their ratio as in Fig.\ref{fig:four}. It is found that at some finite energies, $\Gamma(\omega)$ at  $\alpha=112$ is larger than that at $\alpha=106$, namely,

\begin{eqnarray*}
\lambda(106) \simeq  \lambda(112) \\
\gamma^{\mu}_{\alpha=106} > \gamma^{\mu}_{\alpha=112}\\
\Gamma^{\mu}_{\alpha=106}(\omega_{\rm finite}) < \Gamma^{\mu}_{\alpha=112}(\omega_{\rm finite})
\end{eqnarray*}

This confirms our expectation that the absence of the tight correlation between superconductivity and the low energy spin fluctuation in the first dome is due to the contribution from finite energy spin fluctuation, which itself is not correlated, or in some cases even anti correlated,  with the low energy spin fluctuation.

\begin{figure}[tbp]
\begin{center}
\includegraphics[scale=0.51]{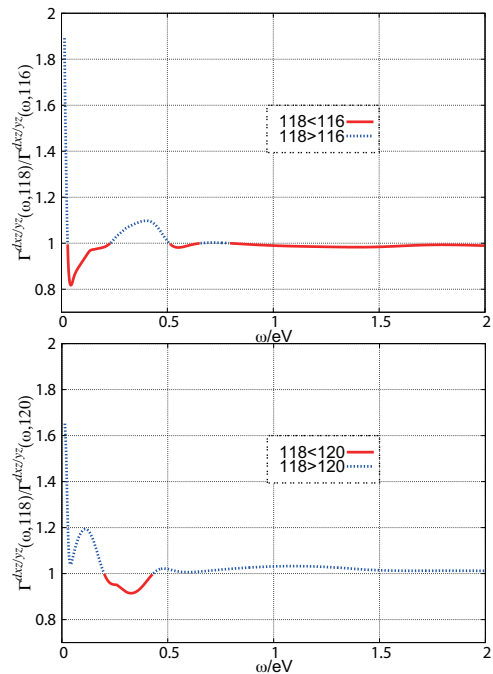}
\caption{Plots similar to Fig.\ref{fig:four}. The ratio of $\Gamma^{xz/yz}$ between $\alpha=116$ and $\alpha=118$ (top), and $\alpha=120$ and $\alpha=118$ (bottom)}
\label{fig:five}
\end{center}
\end{figure}

In the second dome,  on the other hand, the magnitude relation of $\Gamma(\omega)$ among the bond angles $\alpha=116$, 118, 120 depends on $\omega$ as seen in Fig.\ref{fig:five}. The only clear correlation between spin fluctuation and superconductivity is seen at very low frequencies. Therefore, for the second dome, the low energy spin fluctuation seems to dominate the variation of the superconducting eigenvalue $\lambda$.

\section{\label{sec:level4}DISCUSSION}
As described in the Introduction, it was found in the NMR experiments \cite{Mukuda,Ishida} that the correlation between $T_c$ and the development of $1/T_1T$ at low temperatures is material dependent.
In ref.[\onlinecite{Mukuda}] in particular, it has recently been found that the enhancement of $T_c$ in the intermediate regime of the As content in LaFeAs$_{\it x}$P$_{1-{\it x}}$O$_{1-{\it y}}$F$_{\it y}$ is correlated with the development of $1/T_1T$ at low temperatures.
On the other hand, the development of $1/T_1T$ tends to be saturated  upon lowering the temperature in a material (Y$_{\it z}$La$_{1-{\it z}}$FeAsO$_{\it y}$) reaching $T_c=50{\rm K}$, where La is partially substituted  by Y in LaFeAsO$_{\it y}$. 

These experimental results can be understood in terms of the calculation results obtained in the present study by notifying that (i) $\gamma^{xy}$ and $\gamma^{xz/yz}$ mainly contribute to $1/T_1T$, because $xy$, $yz$, $xz$ orbitals are the main origin of the spin fluctuations, and (ii) increasing the phosphorous content corresponds to enlarging the bond angle $\alpha$, while substituting La by Y corresponds to reducing the bond angle. LaFeAsO$_{1-{\it x}}$F$_{\it x}$ has the bond angle of about $\alpha=113\sim 114$, so Y$_{\it z}$La$_{1-{\it z}}$FeAsO$_{\it y}$ is expected to be sitting in the  first dome, where the $d_{xy}$ spin fluctuation plays an important role, and LaFeAs$_{\it x}$P$_{1-{\it x}}$O$_{1-{\it y}}$F$_{\it y}$ is in the second dome, where only the $d_{xz/yz}$ spin fluctuation is important. As was mentioned in the previous section, 
$\gamma$ tends to saturate or even decreases upon lowering the temperature in the right side of the first dome (bond angles $\alpha=111\sim 112$) even though the superconducting $\lambda$ values are substantially larger than those in the second dome. On the other hand, the local enhancement of $\lambda$ is correlated with the development of $\gamma^{d_{xz/yz}}$ at low temperatures in the second dome. 

\begin{figure}[tbp]
\begin{center}
\includegraphics[scale=0.60]{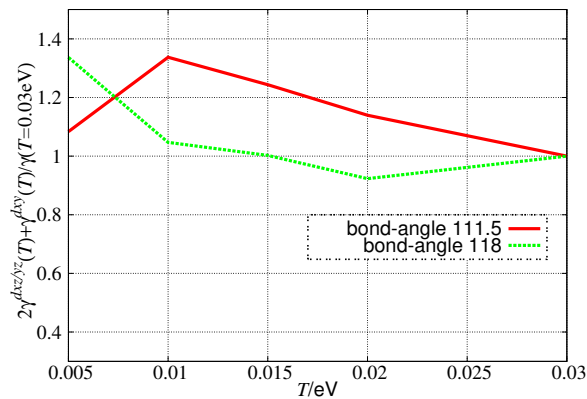}
\caption{$2\gamma^{xz/yz}+\gamma^{xz/yz}$ normalized by its value at $T=0.03$eV, plotted against $T$ for $\alpha=111.5$ and 118.}
\label{fig:six}
\end{center}
\end{figure}

To see more directly the correspondence between the calculation and the NMR experiment, we plot  in Fig.\ref{fig:six} a summation $2\gamma^{xz/yz}+\gamma^{xz/yz}$, which should give the dominant contribution to NMR $1/T_1T$, renormalized by its  value at $T=0.03$eV for angles $\alpha=111.5$ (in the right side of the first dome) and $\alpha=118$ (at which $\lambda$ is locally maximized in the second dome).  Indeed, this quantity starts to decrease for $T<0.01$eV for $\alpha=111.5$, while it continues to develop down to low temperature for $\alpha=118$, in nice correspondence with the experiment in ref.[\onlinecite{Mukuda}]. These results 
strongly indicate that the spin fluctuations originating from multiple orbitals are the origin of the superconductivity in these materials.

\section{\label{sec:level5}CONCLUSION}
In the present work, using the five orbital model of the 1111 iron-based superconductor, we have studied the correlation between the spin fluctuation and the superconducting transition temperature (eigenvalue of the Eliashberg eq.), which exhibits  a double dome feature upon varying the Fe-As-Fe bond angle. Around the first dome with higher $T_c$, the low energy spin fluctuation and $T_c$ are not tightly correlated because the finite energy spin fluctuation also contributes to superconductivity. In fact, even near the $T_c$ maximum, the measure of the low energy spin fluctuation, proportional to $1/T_1T$, tends to saturate or even decrease upon lowering the temperature.
On the other hand, the magnitude of the low-energy spin fluctuation originating from the $d_{xz/yz}$ orbital is correlated with $T_c$ in the second dome.
These calculation results are consistent with recent NMR study\cite{Mukuda}, and  therefore strongly suggest that the pairing in the iron-based superconductors is predominantly caused by the multi-orbital spin fluctuation.

\section*{ACKNOWLEDGEMENT}
We acknowledge stimulating discussions with H. Mukuda, S. Tajima, S. Miyasaka, A. Takemori, K. Tanaka, and K.T. Lai. Numerical calculations were performed
at the facilities of the Supercomputer Center,
Institute for Solid State Physics, 
University of Tokyo. This study has been supported by 
Grants-in-Aid  for Scientific Research No.24340079 and No. 25009605 
from the Japan Society for the Promotion of Science.

\end{document}